# Повышение чувствительности нейтринного телескопа Baikal-GVD с помощью внешних гирлянд оптических модулей.

А. В. Аврорин[1], А. Д. Аврорин[1], В. М. Айнутдинов[1,*], В. А. Аллахвердян[2], З. Бардачова[3], И. А. Белолаптиков[2], И. В. Борина[2], Н. М. Буднев[4], А. Р. Гафаров[4], К. В. Голубков[1], Н. С. Горшков[2], Т. И. Гресь[4], Р. Дворницки[2,3], Ж.-А. М. Джилкибаев[1], В. Я. Дик[2,10], Г. В. Домогацкий[1], А. А. Дорошенко[1], А. Н. Дячок[4], Т. В. Елжов[2], Д. Н. Заборов[1], В. К. Кебкал[5], К. Г. Кебкал[5], В. А. Кожин[6], М. М. Колбин[2], К. В. Конищев[2], А. В. Коробченко[2], А. П. Кошечкин[1], М. В. Круглов[2], М. К. Крюков[1], В. Ф. Кулепов[7], Ю. М. Малышкин[2], М. Б. Миленин[1], Р. Р. Миргазов[4], В. Назари[2], Д. В. Наумов[2], Д. П. Петухов[1], Е. Н. Плисковский[2], М. И. Розанов[8], В. Д. Рушай[2], Е. В. Рябов[4], Г. Б. Сафронов[1], Д. Сеитова[2,10], А. Э. Сиренко[2], А. В. Скурихин[6], А. Г. Соловьев[2], М. Н. Сороковиков[2], А. П. Стромаков[1], О. В. Суворова[1], В. А. Таболенко[4], Б. А. Таращанский[4], Л. Файт[9], А. Хатун[3], Е. В. Храмов[2], Б. А. Шайбонов[2], М. Д. Шелепов[1], Ф. Шимковиц[2,3,9], И. Штекл[9], Э. Эцкерова[3], Ю. В. Яблокова[2]

*[1] Институт ядерных исследований РАН, Москва, Россия, 117312*
*[2] Объединенный институт ядерных исследований, Дубна, Россия, 141980*
*[3] Comenius University, Братислава, Словакия, 81499*
*[4] Иркутский государственный университет, Иркутск, Россия, 664003*
*[5] EvoLogics GmbH, Берлин, Германия, 13355*
*[6] Научно-исследовательский институт ядерной физики им. Д.В. Скобельцына, МГУ, Москва, Россия, 119991*
*[7] Нижегородский государственный технический университет, Нижний Новгород, Россия, 603950*
*[8] Санкт-Петербургский государственный морской технический университет, Санкт-Петербург, Россия, 190008*
*[9] Czech Technical University in Prague, Прага, Чешская Республика, 16000*
*[10] Институт ядерной физики МЭ, Алматы, Республика Казахстан, 050032*

В оз. Байкал продолжается развертывание глубоководного нейтринного телескопа Baikal-GVD. К апрелю 2022 было введено в эксплуатацию 10 кластеров телескопа, в состав которых входит 2880 оптических модулей. Одной из актуальных задач Байкальского проекта является исследование возможностей увеличения эффективности регистрации детектора на основе опыта его эксплуатации и результатов, полученных на других нейтринных телескопах за последние годы. В данной работе рассматривается вариант оптимизации конфигурации телескопа, основанный на установке дополнительной гирлянды оптических модулей между кластерами детектора (внешней гирлянды). Экспериментальная версия внешней гирлянды была установлена в оз. Байкал в апреле 2022 года. В работе представлены результаты расчетов эффективности регистрации нейтринных событий для новой конфигурации установки, техническая реализация системы регистрации и сбора данных внешней гирлянды и первые результаты ее натурных испытаний в оз. Байкал.

*Ключевые слова:* нейтрино, нейтринные телескопы, система сбора данных, Байкал.

* Электронный адрес: *aynutdin@yandex.ru*

**Введение**

Нейтринная астрофизика является новой областью физических исследований. Меньше десяти лет прошло после регистрации первых нейтрино астрофизической природы на установке IceCube. Это, в частности, обусловлено масштабностью детекторов нейтрино высокой энергии – нейтринных телескопов. В настоящее время в мире на разных стадиях развертывания работают три нейтринных телескопа: IceCube на Южном полюсе [1], KM3Net в Средиземном море [2] и Baikal-GVD [3] в оз. Байкал. В настоящее время лидером в области нейтринной астрофизики является нейтринный телескоп IceCube. На нем впервые были выделены кандидаты на нейтрино астрофизической природы в трековом и каскадном режимах регистрации и исследован их энергетический спектр. Вторым по масштабам нейтринным телескопом является Baikal-GVD, крупнейший глубоководный черенковский детектор Северного полушария. Первый полномасштабный кластер Baikal-GVD был введен в эксплуатацию в 2016 году. За последующие 6 лет количество кластеров Baikal-GVD было увеличено до десяти (2880 фотодетекторов - оптических модулей).

В настоящее время одной из актуальных задач Байкальского проекта является исследование возможностей увеличения эффективности регистрации детектора на основе опыта его эксплуатации и результатов, полученных на других нейтринных телескопах за последние годы. Решение этой задачи, в частности, создаст необходимые предпосылки для разработки проекта нейтринного телескопа следующего поколения с эффективным объемом масштаба 10 кубических километров. Именно такой масштаб установки, как показывают эксперименты на нейтринных телескопах, позволит перейти от наблюдения диффузного потока астрофизических нейтрино к исследованию источников их образования. Исследования проводятся в направлениях разработки нового глубоководного фотодетектора (оптического модуля) с увеличенной чувствительной площадью, изучения возможности модернизации системы сбора данных телескопа на базе оптоволоконных линий связи, оптимизации конфигурации регистрирующей системы телескопа. В данной работе рассматривается вариант оптимизации конфигурации телескопа, основанный на установке дополнительных внешних гирлянд оптических модулей между кластерами телескопа в геометрическом центре каждой тройки кластеров. Термин внешняя гирлянда подчеркивает, что она размещается за границей рабочего объема кластера.

Экспериментальная версия внешней гирлянды (ВГ) была установлена в оз. Байкал в апреле 2022 года. В работе представлены результаты расчетов эффективности регистрации нейтринных событий для новой конфигурации установки, техническая реализация системы регистрации и сбора данных внешней гирлянды и первые результаты ее натурных испытаний в составе Байкальского телескопа.

**1. Оптимизация конфигурации Baikal-GVD**

Нейтринный телескоп Baikal-GVD расположен в южной части оз. Байкал. Глубина озера в месте дислокации установки составляет 1366 м. Регистрация излучения в установке Baikal-GVD осуществляется оптическими модулями (ОМ) [4]. В качестве светочувствительного элемента ОМ используется фотоэлектронный умножитель (ФЭУ) Hamamatsu R7081-100. Оптические модули размещаются на гирляндах, установленных на якорях на дне оз. Байкал, и сгруппированы в кластеры. Кластер включает в свой состав центральную гирлянду и семь гирлянд, равномерно расположенных по окружности радиусом 60 метров. Каждая гирлянда состоит из 36 оптических модулей размещенных с шагом 15 метров на глубинах от 750 до 1275 метров. ОМ ориентированы фотокатодами вниз, что повышает эффективность регистрации событий из нижней полусферы и предотвращает потери излучения из-за накопления осадков в верхней части стеклянного глубоководного корпуса оптического модуля.

Оптимизация конфигурации кластеров (расстояния между гирляндами и оптическими модулями) с целью достижения максимальной чувствительности телескопа к потоку астрофизических нейтрино проводилась для нейтринного спектра вида $E^{-2}$. При этих условиях оптимальное расстояние между кластерами составило 300 метров. Энергетический спектр астрофизических нейтрино, в последствии измеренный на установке IceCube, имеет большее значение показателя: $E^{-2.46}$ [5]. С учетом более крутого спектра расстояние между кластерами в 2022 году было уменьшено с 300 м до 250 м и были предприняты меры для увеличения чувствительности межкластерного рабочего объема установки. С точки зрения технической реализации, наиболее эффективным способом повышения чувствительности телескопа является установка дополнительных гирлянд в геометрических центрах каждой тройки кластеров Baikal-GVD – внешних гирлянд (см. рис. 1.1).

Для оценки величины эффекта, связанного с установкой дополнительных внешних гирлянд, моделировался отклик детектора в режимах регистрации мюонов и каскадных ливней, генерированных в нейтринных событиях. Конфигурация регистрирующей системы внешней гирлянды полностью повторяла конфигурацию базовых гирлянд установки. Она включала в себя 36 оптических модулей, расположенных на расстояниях 15 метров по вертикали. Для мюонных событий, имеющих протяженность превышающую геометрические размеры телескопа, эффективная площадь установки для регистрации нейтрино увеличилась пропорционально увеличению количества оптических модулей, что означает отсутствие значимого эффекта от установки внешних гирлянд. Можно отметить только более изотропную чувствительность телескопа, в состав которого включены внешние гирлянды, к потоку окологоризонтальных мюонов.

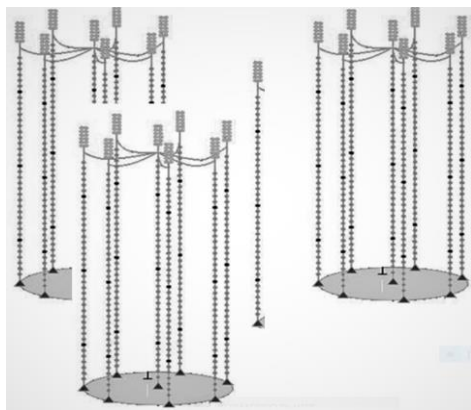

Рис. 1.1. Внешняя гирлянда, расположенная в геометрическом центре трех кластеров Baikal-GVD.

При регистрации каскадов, квазиточечных источников излучения (характерная длина каскадов меньше расстояния между оптическими модулями), ситуация существенно другая. Для оценки эффекта от установки внешней гирлянды было разыграно $10^4$ вершин взаимодействия электронных нейтрино для конфигурации регистрирующей системы, представленной на рис. 1.1. В каждой вершине равномерно разыгрывалось 20 значений косинуса зенитного угла. Для каждого зенитного угла разыгрывалось 20 значений азимутального угла. Для каждого направления было разыграно равномерно 50 значений первичной энергии нейтрино по спектру E$^{-2.46}$ от 1 ТэВ до 10$^5$ ТэВ. Критерием отбора событий являлось ограничение на количество сработавших каналов в кластере N$_{hit}$ >30. Это требование обеспечивает надежное подавление фона от групп атмосферных мюонов. Распределения событий по расстоянию ρ до внешней гирлянды для двух конфигураций телескопа (расстояния между центрами кластеров 250 м и 300 м) представлено на рисунке 1.2.

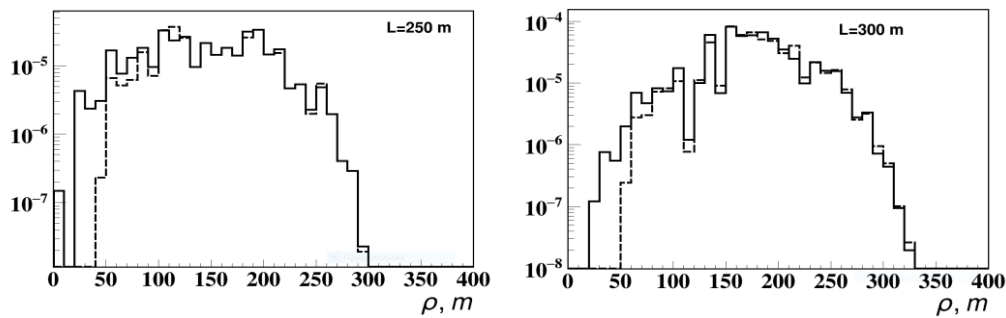

Рис. 1.2. Распределения событий по расстоянию ρ от внешней гирлянды для двух конфигураций: расстояния между центрами кластеров 250 м (слева) и 300 м (справа). Сплошные линии – конфигурация с внешней гирляндой, штриховые – без внешней гирлянды.

При расстоянии между кластерами 250 м, для конфигурации состоящей из 25 гирлянд (24 гирлянды в составе 3-х кластеров и внешняя гирлянда) количество зарегистрированных событий увеличилось на 10%, по сравнению с базовой конфигурацией (24 гирлянды). Для расстояния между кластерами 300 метров рост количества событий составил 5%. Аналогичные расчеты были проведены и для спектра нейтрино вида E$^{-2}$. В этом случае рост количества зарегистрированных событий составил 30% и 10% для расстояний между кластерами 250 м и 300 метров, соответственно. В том случае, если отбирать каскадные ливни с энергией большей 100 ТэВ, из области, где фон событий от атмосферных мюонов и нейтрино становится меньше сигнала от астрофизических нейтрино, рост количества событий составляет 24% для спектра вида E$^{-2,46}$ и расстояния между кластерами 250 м. Полученные результаты показывают значимое увеличение эффективности телескопа в конфигурации, содержащей внешнюю гирлянду. Это инициировало проведение экспериментальных исследований характеристик оптимизированной конфигурации телескопа.

**2. Техническая реализация внешней гирлянды**

Первая внешняя гирлянда была установлена и введена в эксплуатацию в составе нейтринного телескопа Baikal-GVD в апреле 2022 года. К этому моменту времени Baikal-GVD состоял из 10 полномасштабных кластеров. Конфигурация телескопа 2022 г. представлена на рисунке 2.1. Цифры обозначают номера кластеров (нумерация соответствует последовательности их ввода в эксплуатацию), символом *L* обозначены технологические гирлянды, в состав которых входят лазерные калибровочные источники. Многоугольниками отмечено положение внешней гирлянды.

Система регистрации и сбора данных внешней гирлянды в основном идентична базовым гирляндам кластера [6]. Гирлянда состоит из трех секций ОМ (первая секция расположена внизу гирлянды, см. рис. 2.1). В состав каждой секции входит 12 ОМ, до 2-х акустических модемов системы позиционирования [7, 8] и модуль управления секцией. Оптические модули и акустические модемы подключаются к модулю управления секцией, функциями которого являются управление работой ОМ, формирование локального триггера, преобразование аналоговых сигналов ФЭУ в цифровой вид, сбор и первичная обработка данных. Преобразование аналоговых сигналов осуществляется 12-канальным АЦП с частотой дискретизации 200 МГц. На основе данных, поступающих с АЦП, формируются временные кадры событий, содержащие информацию о форме импульсов, регистрируемых на каналах. Управление работой секций, формирование триггера гирлянды и обмен информацией обеспечивает отдельный глубоководный электронный блок – модуль гирлянды (МГ).

Внешняя гирлянда подключена к центру управления 9-го кластера. Центр управления кластера расположен на глубине около 30 метров и связан с Береговой станцией гибридным оптоволоконным кабелем, обеспечивающим электропитание кластера и передачу данных. Подключение внешней гирлянды осуществляется через глубоководный 9-контактный коннектор, ранее предусмотренный для "лазерной гирлянды", предназначенной для межкластерной временной калибровки. В базовой конфигурации кластера коннектор использовался для электропитания и управления лазерными источниками света. При подключении внешней гирлянды через этот коннектор была организована передача дополнительных триггерных сигналов: локального триггера внешней гирлянды (*запрос*) и общего триггера кластера (*подтверждение*).

Данные от секций внешней гирлянды передаются в МГ и от МГ в Центр кластера через удлинители Ethernet IEX-402-SHDSL и затем транслируются на Береговую станцию по оптоволоконной линии связи. Считывание данных с внешней гирлянды осуществляется при поступлении на нее сигнала *подтверждение*, аналогично остальным гирляндам кластера.

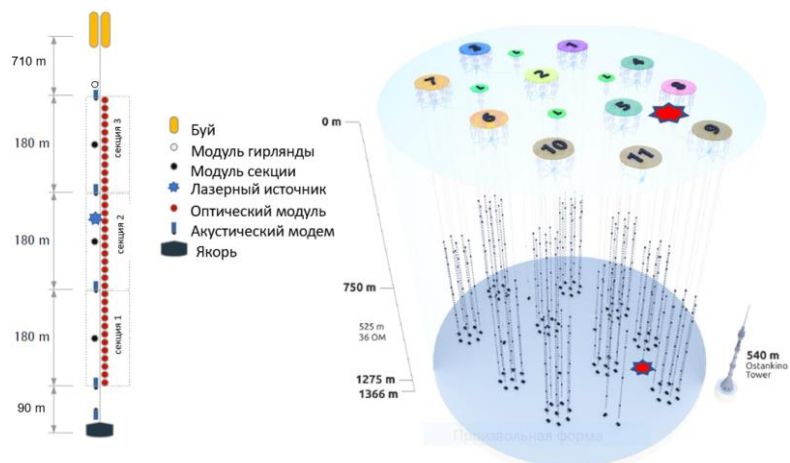

Рис. 2.1. Структура внешней гирлянды и конфигурация Baikal-GVD 2022 года, многоугольником отмечено положение внешней гирлянды в установке.

Единственное отличие внешней гирлянды от базовых гирлянд кластера заключается в том, что в центральной части ВГ размещается лазерный калибровочный источник излучения, подключенный к модулю управления средней секции. Лазерный источник излучает на длине волны 532 нм, энергия в импульсе составляет 0.37 мДж (~$10^{15}$ фотонов) при длительности вспышки ~1 нс. Лазерный источник позволяет осуществлять комплексную оценку эффективности временной и амплитудной калибровки каналов, точности позиционирования ОМ и корректности временной синхронизации кластеров. В состав лазерного источника входит система излучения света, система контроля стабильности интенсивности излучения, управляемый аттенюатор и диффузор, обеспечивающий формирования потока излучения.

Временная калибровка каналов внешней гирлянды осуществляется при помощи светодиодных источников, размещенных в оптических модулях, аналогично базовым гирляндам кластера. Временные сдвижки между каналами обусловлены разностью в длинах кабелей оптических модулей, секций и гирлянд, а также зависимостью временной задержки каналов от высоковольтного напряжения ФЭУ. Калибровочные источники света разработаны на основе светодиодов с длиной волны в максимуме излучения 470 нм и длительностью импульса ~5 нс. Световой импульс формируется в конусе с раствором 15°. В каждом оптическом модуле установлено два калибровочных источника со светодиодами, ориентированными в верхнем направлении и предназначенными для относительной временной калибровки каналов гирлянды. Из-за малого раствора угла их излучения, свет от них не регистрируется фотодетекторами соседних гирлянд. Поэтому для определения временных сдвижек между каналами внешней гирлянды и гирляндами окружающих ее кластеров, используются две матрицы светодиодов, которые устанавливаются в оптические модули средней и верхней секции ВГ. Матрицы состоят из 10 светодиодных источников, ориентированных горизонтально и расположенных равномерно по окружности.

### 3. Натурные исследования внешней гирлянды

Точность восстановления геометрии событий (направления движения мюонов, положения оси каскадов и вершины их взаимодействия) в нейтринном телескопе зависит от точности измерения времени регистрации сигналов, поступающих с ФЭУ. Ошибки, возникающие при измерении времени регистрации, определяются двумя факторами: точностью измерения относительных временных сдвижек каналов (временной калибровкой) и неопределенностью времен регистрации сигналов (временной синхронизацией). Натурные испытания, проведенные в 2022 году, кроме долговременных исследований корректности работы аппаратуры внешней гирлянды, включали в себя и исследования точности ее временной калибровки и синхронизации с другими гирляндами телескопа.

3.1. Временная калибровка внешней гирлянды

Аппаратура системы временной калибровки телескопа была разработана для базового варианта Baikal-GVD, для которого расстояниями между гирляндами составляет 60 метров. При этом условии точность калибровки каналов составляет величину ~2 нс. При увеличении расстояния от калибровочного источника до ~80 м, при длине поглощения света в байкальской воде 22 м величина калибровочного сигнала уменьшается более чем в пять раз, что сказывается на точности временной калибровки внешней гирлянды. Для исследования этой точности в период с апреля по сентябрь 2022 г. было проведено несколько калибровочных серий измерений. В качестве калибровочных источников использовались светодиодные матрицы, установленные на внешней гирлянде. Пример калибровочного события, зарегистрированного на трех кластерах представлен на рис. 3.1.1. Кружками выделены сработавшие каналы, цвет показывает время регистрации сигналов на каналах. Нулевой отсчет времени находится в середине временного кадра АЦП, длительностью 5 мкс и соответствует моменту регистрации сигнала *подтверждение* аппаратурой модуля секции, инициированного *запросом* этой секции. Выбор шкалы времени осуществляется на этапе настройки установки при помощи программируемых задержек времен формирования кадров событий.

Для вычисления относительных временных сдвижек между каналами внешней гирлянды и базовых гирлянд кластеров определялась разность $\Delta T$ между ожидаемыми из геометрии $dT_0$ и измеренными $dT$ задержками между сигналами, зарегистрированными на этих гирляндах. Каналы в пределах каждой гирлянды калибровались при помощи вертикальных светодиодов, встроенных в каждый ОМ (см. п. 2). От вспышки матрицы светодиодов срабатывает несколько каналов на каждой гирлянде. Это позволяет оценить точность калибровки, которая определяется разбросом измеренных для этих каналов значений $\Delta T$. Для того чтобы минимизировать вклад в ошибку измерений неопределенности в определении положения ОМ акустической системой позиционирования (~30 см), разность значений $\Delta T$ определялась для пар соседних каналов на одной гирлянде. Точность калибровки характеризовалась средним абсолютным значением параметра $\tau = (\Delta T_i - \Delta T_{i+1})$, где $i$ – номер канала. На рисунке 3.1.2 показаны примеры распределения калибровочных событий по величине параметра $\tau$, измеренного для внешней гирлянды относительно гирлянд кластеров 5, 8 и 9.

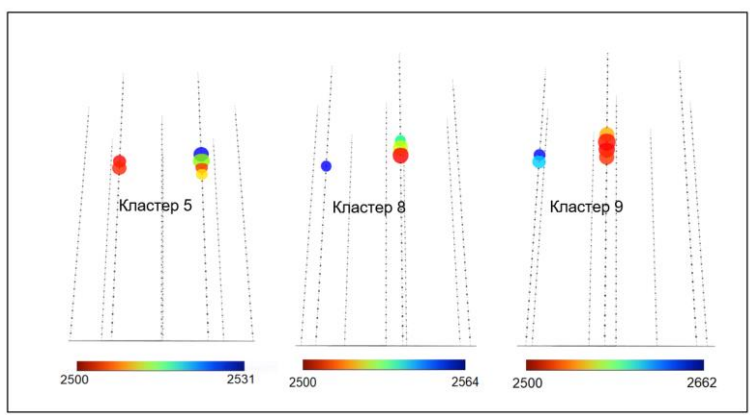

Рис. 3.1.1. Вид калибровочного события от матрицы внешней гирлянды на 3-х окружающих ее кластерах телескопа в проекциях на источник излучения. Шкала времени регистрации представлена под каждым рисунком в наносекундах.

Пример результатов временной калибровки внешней гирлянды относительно окружающих ее кластеров для одной серии калибровочных измерений (4.05.2022) представлен в таблице 3.1. В таблице представлено расстояние $R$ от внешней гирлянды до ближайших базовых гирлянд окружающих ее кластеров, количество каналов $K$ базовых гирлянд, по которым проводилась временная калибровка, средний заряд $Q$ на каналах, измеренный в фотоэлектронах (ф.э.), и средняя величина абсолютных значений параметра $\tau$, вычисленная по парам каналов гирлянд.

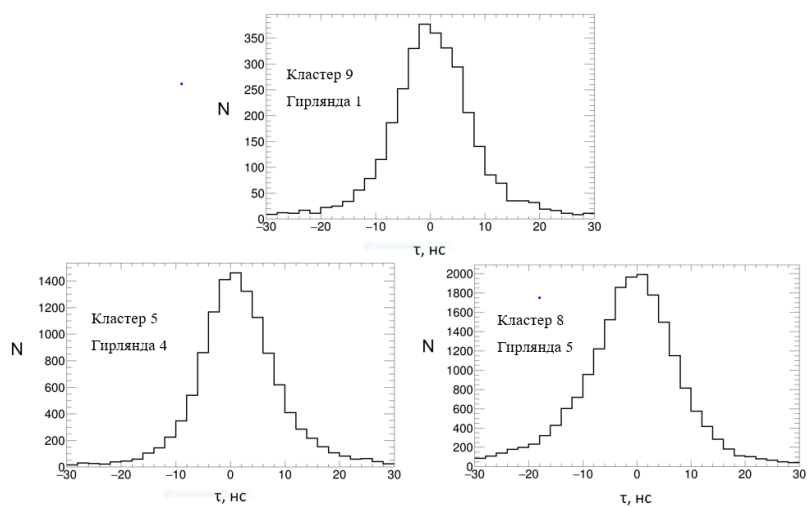

Рис. 3.1.2. Примеры распределения событий по величине параметра $\tau$ в режиме калибровки внешней гирлянды относительно кластеров 5, 8 и 9.

Таблица 3.1. Калибровка внешней гирлянды относительно кластеров телескопа.

| Номер кластера | $R$, м | $K$ | $Q$, ф.э. | $\tau$, нс |
|---|---|---|---|---|
| Кластер 5 | 85 | 5 | 3,6 | 2,8 |
| Кластер 8 | 80 | 2 | 4,6 | 0,4 |
| Кластер 9 | 77 | 3 | 3,5 | 2,5 |

Неопределенность временной калибровки внешней гирлянды относительно кластеров телескопа составляет величину меньше 3 нс, близкую к значению, полученному при калибровке базовых гирлянд телескопа, и является приемлемой с точки решения задачи восстановления геометрии физических событий в телескопе [9]. Величина неопределенности калибровки определяется рассеянием света в воде, которое при расстояниях от источника ~80 метров становится существенным и не учитывалось при расчете времен срабатывания каналов. Заряды сигналов на калибровочных каналах варьируются от канала к каналу в пределах от 2 ф.э. до 10 ф.э., в среднем около 4 ф.э. Следует отметить, что величина заряда слабо коррелирует с расстоянием от внешней гирлянды до кластера. Это связано с тем, что для светодиодов калибровочных матриц не проводился специальный отбор по максимальной величине излучаемого светового импульса. Значение максимального сигнала для разных экземпляров светодиодов может значительно различаться (в 3 – 4 раза), что нарушает изотропность светового потока от калибровочной матрицы.

### 3.2. Временная синхронизация

Система временной синхронизации Baikal-GVD обеспечивает работу всех каналов установки в условиях единой шкалы времени. Она включает в себя две подсистемы, обеспечивающие синхронизацию каналов в пределах одного кластера, и синхронизацию кластеров друг с другом [9]. Работа этих подсистем основана на разных принципах. Синхронизация каналов в пределах одного кластера осуществляется при помощи общего триггера, формируемого в Центре управления кластера и транслируемого на все его секции. Для синхронизации работы кластеров на каждом из них измеряется время формирования этого общего триггера. Измерение осуществляется двумя независимыми системами: WR (White Rabbit) [10] и SSBT (Synchronization System of Baikal Telescope) [9], специально разработанной для Baikal-GVD.

Для исследований работы системы синхронизации внешней гирлянды был проведен анализ данных калибровочных серий измерений в режиме одновременной засветки гирлянды и окружающих ее кластеров лазерным источником света, расположенном на ВГ. В качестве параметра, характеризующего точность синхронизации, использовалась величина среднеквадратичного отклонения (*СКО*) измеренной задержки времен срабатываний пар синхронизуемых каналов - *dt*. Для анализа отбирались каналы, для которых средний заряд сигналов от лазера превышал 10 фотоэлектронов. Это позволило минимизировать влияние статистической ошибки измерения времен регистрации сигналов.

Рисунок 3.2.1 иллюстрирует точность синхронизации гирлянды с окружающими ее кластерами 5, 8 и 9. Следует подчеркнуть, что синхронизация ВГ с кластером 9, в состав которого она входит как 9-я гирлянда, осуществлялась по общему триггеру кластера, в то время как для кластеров 5 и 8 дополнительно измерялось время между триггерами кластеров при помощи системы WR. Точность временной синхронизации внешней гирлянды (СКО) с окружающими ее кластерами составила величину 2,1 – 2,2 нс, что хорошо согласуется с ожидаемым значением 2,0 нс, которое определяется дискретностью времен фиксации триггерного сигнала аппаратурой секций установки, равной 5 нс.

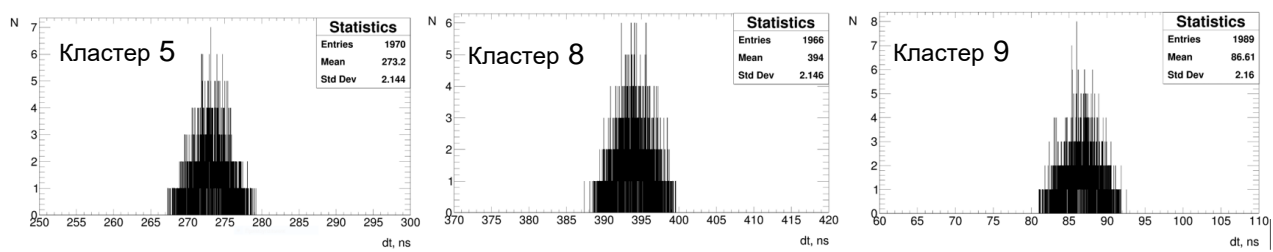

Рис. 3.2.1. Распределение событий по разнице времен регистрации сигналов *dt*, на внешней гирлянде и на ближайших к ней гирляндах кластеров 5, 8 и 9.

3.3. Работа внешней гирлянды в режиме регистрации событий

В апреле 2022 года внешняя гирлянда была введена в эксплуатацию в составе нейтринного телескопа Baikal-GVD в режиме постоянной экспозиции. Анализ полученных с внешней гирлянды экспериментальных данных показал корректность ее работы и не выявил особенностей ее функционирования по сравнению с остальными гирляндами телескопа. В качестве иллюстрации на рисунке 3.3.1 показана частота формирования локальных триггеров секциями внешней гирлянды по сравнению с гирляндой 4 кластера 5, расположенной в 85 метрах от нее. Условием формирования триггера секции является совпадений сигналов с двух ее соседних оптических модулей во временном окне 100 нс, превышающих пороговые значения 1,5 ф.э. и 4,0 ф.э.

Скорость счета локальных триггеров определяется тремя основными процессами: регистрацией атмосферных мюонов, собственными шумами ФЭУ и фоновым свечением воды оз. Байкал. Временная зависимость и абсолютная величина темпов счета, измеренных на двух гирляндах, находятся в хорошем согласии. На графиках рис. 3.3.1 отражены основные особенности функционирования секций оптических модулей телескопа: рост числа срабатывания с уменьшением глубины размещения ОМ, связанный с увеличением вклада фонового свечения воды; стабильное поведение скорости счета в весенне-зимний период и его значительные флуктуации в летний и осенний сезоны, обусловленные переносом светящихся органических масс течениями озера (хемолюминисценция). В период нестабильного светового фона озера темп счета секций может достигать нескольких десятков Герц. Это приводит к повышению частоты фоновых срабатываний кластера до уровня, превышающего возможности системы передачи данных. Поэтому, во избежание потери данных, в этот период пороги регистрации каналов искусственно поднимаются на 20 – 30 %. Эффект повышения порогов регистрации в августе 2022 года можно наблюдать на рис. 3.3.1 как снижение темпа счета *запросов* для первых (нижних) секций гирлянд.

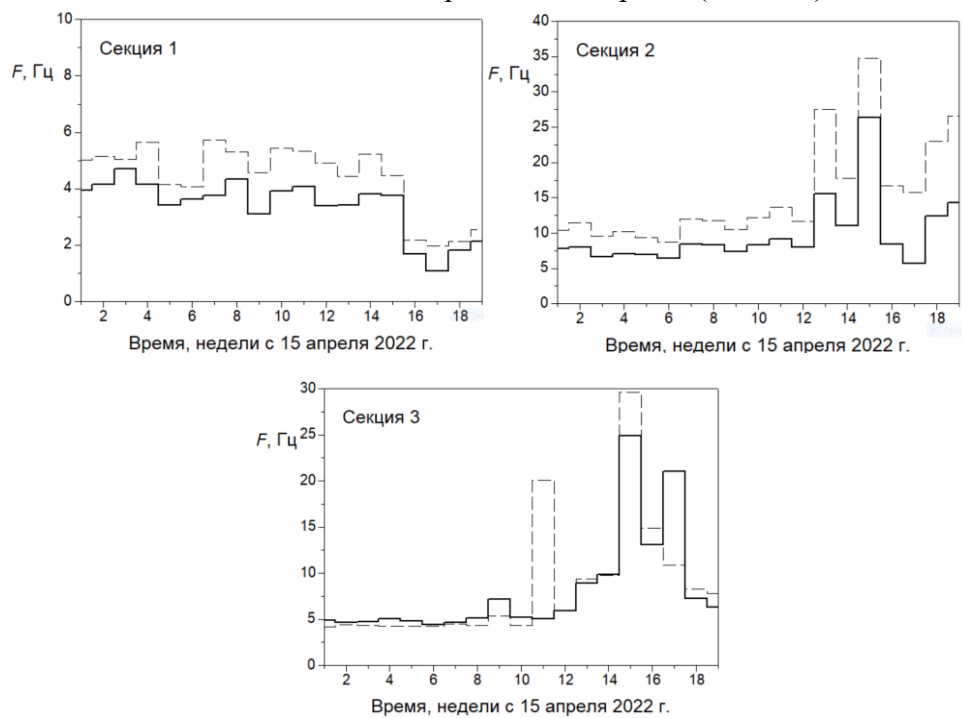

Рис. 3.3.1. Частота локальных триггеров, формируемых нижней (№1), средней (№2) и верхней (№3) секциями внешней гирлянды (сплошная линия) и гирлянды 4 кластера 5 (штриховая линия).

**Заключение**

Представленные в статье расчеты показывают значимое увеличение эффективности работы регистрирующей системы Baikal-GVD при установке внешней гирлянды между кластерами телескопа. Для конфигурации из трех кластеров (24 гирлянды) добавление одной внешней гирлянды обеспечивает увеличение количества регистрируемых событий на 10% и 24% для каскадных ливней с энергией большей 1 ТэВ и 100 ТэВ, соответственно. Техническая реализация проекта по установке внешних гирлянд в геометрических центрах каждой тройки кластеров не представляет принципиальных затруднений. Система сбора данных, глубоководная кабельная инфраструктура и система электропитания кластера могут быть достаточно просто адаптированы для обслуживаний 9-и гирлянд (включая одну внешнюю) вместо 8-и.

Натурные испытания внешней гирлянды показали корректность и надежность функционирования входящей в ее состав аппаратуры, и достаточно высокую точность ее временной калибровки относительно окружающих гирлянд (~2,5 нс). Точность синхронизации внешней гирлянды с кластерами телескопа составила величину ~2 нс, совпадающую с точностью синхронизации базовых гирлянд телескопа. Использование светодиодной калибровочной системы внешней гирлянды позволило провести относительную временную калибровку кластеров, что обеспечивает дублирование лазерной калибровочной системы телескопа, которая в настоящее время используется для этих целей. 

**Список литературы**

# Increasing the sensitivity of the Baikal-GVD neutrino telescope by using external strings of optical modules.

© 2022  V. A. Allakhverdyan[1], A. D. Avrorin[2], A. V. Avrorin[2], V.M. Aynutdinov[2*], Z. Bardačová[3], I. A. Belolaptikov[1], I. V. Borina[1], N. M. Budnev[4], V. Y. Dik[1,10], G. V. Domogatsky[2], A. A. Doroshenko[2], R. Dvornický[1,3], A. N. Dyachok[4], Zh.-A. M. Dzhilkibaev[2], E. Eckerová[3], T. V. Elzhov[1], L. Fajt[5], A. R. Gafarov[4], K. V. Golubkov[2], N. S. Gorshkov[1], T. I. Gress[4], K. G. Kebkal[6], V. K. Kebkal[6], A. Khatun[3], E. V. Khramov[1], M. M. Kolbin[1], K. V. Konischev[1], A. V. Korobchenko[1], A. P. Koshechkin[2], V. A. Kozhin[7], M. V. Kruglov[1], M. K. Kryukov[2], V. F. Kulepov[8], Y. M. Malyshkin[1], M. B. Milenin[2], R. R. Mirgazov[4], D. V. Naumov[1], V. Nazari[1], D. P. Petukhov[2], E. N. Pliskovsky[1], M. I. Rozanov[9], V. D. Rushay[1], E. V. Ryabov[4], G. B. Safronov[2], D. Seitova[1,10], B. A. Shaybonov[1], M. D. Shelepov[2], F. Šimkovic[1,3,5], A. E. Sirenko[1], A. V. Skurikhin[7], A. G. Solovjev[1], M. N.Sorokovikov[1], I. Štekl[5], A. P. Stromakov[2], O. V. Suvorova[2], V. A. Tabolenko[4], B. A. Tarashansky[4], Y. V. Yablokova[1], D. N. Zaborov[2]

[1] *Joint Institute for Nuclear Research, Dubna, Russia, 141980*

[2] *Institute for Nuclear Research, Russian Academy of Sciences, Moscow, Russia, 117312*

[3] *Comenius University, Bratislava, Slovakia, 81499*

[4] *Irkutsk State University, Irkutsk, Russia, 664003*

[5] *Czech Technical University in Prague, Prague, Czech Republic, 16000*

[6] *EvoLogics GmbH, Berlin, Germany, 13355* 12

[7] *Skobeltsyn Institute of Nuclear Physics MSU, Moscow, Russia, 119991*

[8] *Nizhny Novgorod State Technical University, Nizhny Novgorod, Russia, 603950*

[9] *St. Petersburg State Marine Technical University, St. Petersburg, Russia, 190008*

[10] *Institute of Nuclear Physics ME RK, Almaty, Kazakhstan, 050032*

The deployment of the Baikal-GVD deep underwater neutrino telescope is continuing in Lake Baikal. By April 2022, ten clusters of the telescope were put into operation, with 2880 optical modules in total. One of the relevant tasks in this context is to study the possibilities of increasing the efficiency of the detector based on the experience of its operation and the results obtained at other neutrino telescopes in recent years. In this paper, a variant of optimizing the configuration of the telescope is considered, based on the installation of additional strings of optical modules between the clusters (external strings). An experimental version of the external string was installed in Lake Baikal in April 2022. This paper presents a first estimate of the impact of adding external strings on the neutrino detection efficiency, as well as the technical implementation of the detection and data acquisition systems of the external string and first results of its in-situ tests.

*E-mail: aynutdin@yandex.ru